# ESR evidence for disordered magnetic phase from ultra-small carbon nanotubes embedded in zeolite nanochannels

S. S. RAO[1]*, A. STESMANS[1], J. V. NOYEN[2], P. JACOBS[2], and B. SELS[2]

[1] Department of Physics and INPAC-Institute for Nanoscale Physics and Chemistry, University of Leuven, Celestijnenlaan 200 D, B-3001 Leuven, Belgium

[2] Department of M²S, Centre for Surface Science and Catalysis, KU Leuven, Kasteelpark Arenberg 23, 3001 Heverlee, Belgium.

**Abstract -** A multi-frequency electron spin resonance (ESR) study provides evidence for the occurrence of low temperature ferromagnetic/spin-glass behavior in aligned arrays of sub-nanometer single walled carbon nanotubes confined in zeolite nano-channels, owing to $sp^2$-type non-bonding carbon associated localized states with density of $\approx 3 \times 10^{19}$ /g. Features related to the much anticipated conduction ESR are not detected. In the paramagnetic phase, the ESR linewidth is found to be weakly dependent on microwave frequency.

*Corresponding author. Fax**:** + 32 16 32 79 87. E-mail address: Srinivasarao.singamaneni@fys.kuleuven.be (S S Rao)

**Introduction. -** Impressive research efforts have been devoted to study the properties of carbon nanotubes (CNTs) to incorporate them in electrical devices as interconnectors and in transistors [1, 2]. CNTs are also promising for spin-based applications such as spin-qubits and spintronics because the weak spin-orbit (SO) coupling would make it possible to transfer information over long distances [3]. In addition to their projected potential applications, much attention has been drawn to novel physics involved, such as the appearance of the Luttinger Liquid state [4], prediction of Aharonov-Bohm oscillations of the band gap [5], and low temperature superconductivity [6] – to name a few.

There has been much progress in the investigation of the uncommon non metal related Π-electron magnetism in the carbon based nanoscale materials. The nanoscale carbon materials, such as single walled carbon nanotubes (SWCNTs), activated carbon fibers (ACFs), proton irradiated graphite and rhombohedral $C_{60}$ (Rh-$C_{60}$), carbon nano-foam, and carbon nanoclusters, were shown to exhibit several interesting magnetic properties, like anti-ferromagnetic, ferromagnetic, ferri-magnetic and spin-glass features [7-10]. An intercalation-induced spin-glass feature has been observed in nano-crystalline diamond, attributed to $sp^3$ bonding [11]. The origin of magnetism observed in carbon has variably been ascribed to dangling bonds (DBs) associated with defects, states localized on edge carbon atoms, the possible role of hydrogen, negative Gaussian curvature and metallic ferromagnetic impurities. Nevertheless, recent work [12] has provided interesting direct experimental evidence for the intrinsic magnetic nature of carbon. It has been reported [13] that there exists a clear correlation between the presence of defects and observed ferromagnetism in carbon based materials. All the experimental results reported so far were obtained on bundles of interacting carbon materials of 'large' diameters.

To probe the intrinsic physical properties of isolated non-interacting individual CNTs, Tang *et al.* [14] have fabricated SWCNTs of 0.4 nm diameter –the smallest so far- inside the nanochannels of porous zeolite $AlPO_4$-5 (AFI) single crystals. These are formed after the pyrolysis of tripropylamine (TPA) molecules [6], and are considered to be ideal one-dimensional quantum hollow wires. The electronic and magnetic properties of these ultra-small CNTs are expected to be drastically different from those of large-sized CNTs because of the strong admixing of σ-Π unoccupied electronic orbitals due to curvature effects. Quite interestingly, the former tubes were shown to exhibit unusual novel phenomena like diamagnetism and superconductivity at low temperatures [6] in addition to several other optical properties and doping induced effects [15].

Electron spin resonance (ESR) spectroscopy is a most suitable local probe to investigate the nature and dynamics of magnetic species. Over the previous decade or so, ESR work has been performed on several forms of carbon such as an ensemble of large diameter single and multi-walled carbon nanotubes, carbon nanoclusters and shell-like particles. Particularly, diverse ESR results have been reported on pristine large diameter bundles of CNTs, intercalated and doped CNTs, and irradiated CNTs manufactured by various methods, aiming to investigate distinct magnetic phases present in these materials. Most prominently, three different types of signals are observed: a broad featureless ESR signal centered at g value $g_c$=2 [16], originating from residual metallic catalytic particles; a second signal at g = 2.07 [17] ascribed to the conduction electron spin resonance (CESR) of delocalized itinerant spins; a third narrower signal (g = 2.00) [18] is assumed to stem from the localized electron spins. Here, we should add that while several research groups have failed to observe the CESR [19], a few have claimed to detect the asymmetric Dysonian-shaped CESR signal from pristine SWCNT bundles [7]. Recent theoretical ESR studies [20] predict that SWCNTs are expected to show a two peak ESR spectrum (fine structure) due to spin-charge separation, while MWCNTs are expected to exhibit an asymmetric ESR signal due to strong Rashba type SO coupling. To observe these features experimentally, it is suggested to work with clean systems, free from metallic catalytic impurities, comprised of isolated carbon nanotubes with no inter-tube interaction, and at sub-one Kelvin temperatures.

To date, it appears no attention has been paid to investigate the nature and dynamics of magnetic species in confined sub-nanometer metal free and aligned CNTs formed within a zeolite. According to previous suggestions though, it would constitute inherently a much appropriate system to verify theoretical predictions. This concerns the subject of the present work where the nature of magnetic species in clean ultra–small CNTs grown in nano channels of zeolites is probed by multi frequency ESR. We report on evidence for the occurrence of a ferromagnetic/spin-glass phase at low temperatures, the origin of which is found to be due to the presence of non-bonding localized carbon bonds. No evidence is found for the much sought CESR signal for none of the three microwave bands used.

**Experimental. -** The sample preparation method [21,22] involves heat treatment of a SAPO 5 in an inert atmosphere (pyrolysis) and filling its pores with a suitable carbon source. The approach results in the presence of only three chiralities (5,0) (4,2) and (3,3), thus minimizing chiral distribution. The obtained pore filling factor is ~21% of the theoretical value, which is significantly higher than the values reported in other works [23]. X-ray diffraction studies indicate the crystalline nature of nanotubes, where it has also been inferred that (5,0) and (3,3) chiral tubes are metallic, and (4,2) tubes are semiconducting. Raman radial breathing mode (RBM) features indicate an average inner diameter of 0.4 nm for the SWCNTs. From the optical polarized photoluminescence data, the arrays of CNTs are found to align according to the channels of the zeolite crystal. The ESR samples studied, designated as SWCNT@SAPO 5, imply that single walled carbon nanotubes (SWCNT) are occluded inside the channels of a non-magnetic insulating SAPO 5 zeolite crystal. For reasons of comparison, ESR observations have also been carried out on free standing SWCNTs obtained through dissolution of the zeolite matrix in aqueous acidic solution. The specific method aims to produce sub-nanometer CNTs with high tube (filling) densities, with narrow diameter distribution, and of limited number of chiralities. The approach has been shown to enable excellent control on the tube diameter and chiralities. Importantly, the presence of metal is not needed to grow the nanotubes. Also, the growth occurs below 500 $^o$C, lower than temperatures ($\geq$750 $^o$C) reported previously, and hence this growth technique is believed to be compatible with most microelectronic technology platforms.

As described elsewhere [24], continuous wave ESR experiments have been carried out in the range 1.5 – 90 K using home–built K-band ($\approx$ 20.6 GHz) and X-band ($\approx$ 8.9 GHz) set ups and a Q-band ($\approx$ 34 GHz) Bruker EMX spectrometer − all operated under conditions of adiabatic slow passage. Conventional low power first-derivative-absorption $dP_{\mu r}/dB$ ($P_{\mu r}$ being the reflected microwave power) spectra were detected through applying sinusoidal modulation ($\approx$100 kHz, amplitude $B_m \approx$ 0.3G) of the externally applied magnetic field $\vec{B}$, with incident microwave power $P_\mu$ as well as $B_m$ cautiously reduced to avoid signal distortion. The spin density was quantified by double numerical integration of the K-band derivative absorption spectra, and making use of a co-mounted calibrated Si:P intensity marker with g(1.5 K = 1.99876) [24], which also served as g marker. Obtained relative and absolute accuracies on defect densities are estimated at ≈ 6% and 15%, respectively.

**Results and Discussion. -** Figure1(a) shows low-power K-band ESR spectra measured on SWCNT@SAPO 5 at 1.5, 20, 50, and 80 K. At all the temperatures covered, a symmetric isotropic ESR signal is observed at zero–crossing g value $g_c$ ≈ 2.0025, with corresponding spin density ≈ 3 × 10$^{19}$ g$^{-1}$. This g value falls within the range of reported carbon ESR signals (g = 2.0022 – 2.0035), and may be ascribed to C-related dangling bonds of spin S = ½. No other ESR signal could be detected over a broad magnetic field sweep range up to 9000 G (cf. Fig. 1(b)). The line shape factor (κ) is calculated using the formula κ ≡ I/(h×ΔB$^2_{PP}$), with ΔB$_{PP}$ the peak–to-peak signal width, with h the peak-to-peak height and I is the doubly integrated $dP_{\mu r}/dB$ signal. The determined line shape factors over the covered temperatures (1.5 – 90 K) fall in the range of 1.08 – 1.58, close to the value κ$^G$ = 1.03, for a Gaussian line. ESR spectral parameters (cf. Fig.2) are inferred at various T's by means of the computer simulations generally pointing to Voigt line shapes of predominant Gaussian character.

Figure 2 depicts the temperature dependence of ESR spectral parameters for SWCNT@SAPO 5, revealing several noteworthy aspects. The T-dependence of the ESR line width can be adequately described as $\Delta B_{pp}(G) = \Delta B_{pp}(\infty) + A\exp(-T/T_g)$, where $\Delta B_{PP}(\infty)$ represents the high-T linewidth and A is proportionality constant, and $T_g$ is the spin glass transition temperature. Optimized fitting (continuous line in Fig. 2(a)) results in $T_g = 7 \pm 2$ K. Thus, as shown in Fig. 2(a), a sharp rise in peak-to-peak width $\Delta B_{PP}$ is observed for temperatures decreasing below $\sim 2T_g$ while it is almost constant at high T ($T>2T_g$), which is a characteristic feature of a spin-glass/ferromagnetic system [25]. Unlike other phase transitions, one does not see a divergence in linewidth at the spin-glass transition. The observed line broadening can be attributed either to the enhancement of the spin-relaxation rate on approaching $T_g$ from the high temperature side or to a distribution of local internal fields. Figure 2(b) shows a plot of g versus T, indicating $g_c$ to be independent of T. As observed on the CNTs in zeolite cages, the ESR signal exhibits a predominantly Gaussian character at 8.9 GHz (4.2 K), with a $\Delta B_{pp} \sim 8.3$ G and g ~ 2.00251; For the free standing SWCNTs, the ESR signal takes a more Lorentzian shape at 8.9 GHz (4.2 K), now with $\Delta B_{pp} \sim 5$ G and with g ~ 2.00254. So, from the above observed distinct changes in the ESR spectral properties of CNTs with and without cages, it can be inferred that the ESR signal indeed stems from the CNTs present inside the nanochannels of the zeolite.

The average g value lies below 2.003, which is close to that of a carbon DB defect of spin S = 1/2. This has been considered [26] as a characteristic feature of $sp^2$ C hybridization, in which the unpaired spin lies within the curved layer structure of a 0.4 nm diameter SWCNT − a wrapped sheet of graphene. Very recently [27], Kuemmeth *et al.* have measured a g=2.07 signal on single CN quantum dots using tunneling spectroscopy, which they attributed to itinerant CE spins, the large g shift arising from curvature induced spin-orbit coupling. The currently observed g value ($\approx$ 2.00251) is inconsistent with the earlier measured values (g=2.05-2.07) [28] of CESR on non-embedded single and multi walled CNTs of larger diameter ($\geq 20 - 50$ nm). To gain further insight and check for possible signatures of CESR, the T dependence of the ESR signal strength (area under the absorption curve $\propto$ to number of spins) is plotted in Fig. 2(c). The exposed T dependences allows us to consider the signal as originating from a system comprised of C-related paramagnetic centers with strongly localized states. As evident from the $1/\chi_{ESR}$-vs-T plot (cf. inset in Fig.2(c)), the system susceptibility $\chi_{ESR}$ largely follows a Curie-Weiss behaviour ($\chi_{ESR} \propto \dfrac{C}{T-T_c}$, where C is Curie-Weiss constant). The Curie-Weiss temperature ($T_C$) is inferred as $4 \pm 1$ K, close to the $T_g$ value (7 K) obtained from the T dependence of the ESR linewidth data.

The appearance of a broadened and intense ESR signal at low temperatures suggests that at least some of the unpaired spins are ferromagnetically correlated (short-range coupling). The revealed T-dependent properties of ESR spectral parameters provide strong evidence for the spin-glass/ferromagnetic characteristic nature of the originating C-related spin system. The apparent absence of resolved H-related ESR hyperfine features as well as non-carbon related ESR signal(s) would counter extrinsic impurities as origin of the observed magnetism. As anticipated for a CESR signal, one would expect to observe Pauli-type spin susceptibility, an increase in line width with T, and a large g shift from that of free electron value (2.00232) due to curvature induced SO-coupling. Apparently, any signature of such long sought itinerant (delocalized) conduction electron spin system is absent.

To probe further, additional ESR measurements were performed at two other distinct microwave frequencies, $\approx 8.9$ and $\approx 34$ GHz, and at various temperatures. At all temperatures, a single isotropic symmetric ESR signal is detected, of which computer assisted analysis leads to results (not shown) consistent with those of the K-band data. We also find that g value is independent of frequency throughout the measured temperature range. In the paramagnetic phase (at 70 K), the ESR signal width $\Delta B_{PP}$ is found to be weakly dependent on microwave frequency as shown in Fig. 3. From this figure, we obtain the frequency dependent $\Delta B^{f}_{pp} = 0.018$ G/GHz, a Gaussian part. This can be attributed to a g-distribution effect resulting from different local environments of SWCNTs. But it is only a minor contribution. The frequency independent part can be due to dipolar interaction, unresolved hyperfine interaction –perhaps, arising from hydrogen nuclei - resulting in inhomogeneous Gaussian broadening and also contribution from molecular oxygen.

We next address the possible origin of spin-glass/ferromagnetic phase. It has been reported that direct interaction (Heisenberg-type) between the localized spins with [8] or without [29] mediation of itinerant spins can give rise to ferromagnetism in proton-irradiated graphite. Relevant also is that the observed density of localized spins, $\approx 3 \times 10^{19}$ spins/g, was shown sufficient to give rise to an ordered/disordered magnetic phase in carbon based systems [30]. The fact that the current ESR signal is observed down to the temperature of 1.5 K (K-band) rules out the appearance of anti-ferromagnetic and diamagnetic features. It has been shown both experimentally and theoretically that the single layer graphene, graphene nanoribbons, and CNTs contain a mixture of $sp^2$ and $sp^3$ hybridization [26]. An explicit evidence for the presence of $sp^2$ hybridized C comes from our measured average g value, which is less than 2.003. The presence of an, albeit limited, g distribution as inferred from the Gaussian part in the ESR lineshape and the weak linear increase of $\Delta B_{PP}$ with frequency, indicates the presence of different types of spin centers, probably related with $sp^2$ and $sp^3$ hybridization.

On the basis of above results, the origin for the appearance of the spin-glass feature, a disordered magnetic phase, can be understood as a result of the strong competition between two different magnetic spin regions, a dominant set of $sp^2$ regions besides a minority phase of $sp^3$ character [26]. This frustrated magnetism translates into a form of spin-glass phase as it is quite easy to nucleate magnetic moments in a defective $sp^2$ carbon network. Element-specific X-ray magnetic circular dichroism (XMCD) measurements [12] on proton irradiated nanoscale carbon films firmly established that the origin of ferromagnetism in carbon is due to spin-polarized $\Pi$ electrons of carbon. In addition to localized electronic states, flat-band states with topological line defects [31], the interplay between the dimensionality, and the curvature effects of the nanotubes can also give rise to the observed ferromagnetic spin ordering.

To gain more insight and to investigate the possible role of hydrogen in the occuring magnetic properties, we subjected same samples to hydrogen treatment in pure $H_2$ (99.9999 %; 1 atm, 500 °C) for $\approx$ 180 min. This turned out to be unsuccessful in passivating the C–related defects, leaving unaffected the low temperature magnetic features. Still no CESR could be detected. As possible reason for the absence of CESR from SWCNTs grown in zeolite nanochannels, we speculate that either the Luttinger liquid phase broadens the CESR beyond the ESR detection limit [32] or the itinerant spins are trapped at defect states, essentially meaning that the number of metallic nanotubes present is insufficient. Clearer conclusion has to await further work.

**Conclusions.** - We have presented first multi-frequency electron spin resonance measurements on ultra-small single walled carbon nanotubes embedded in a SAPO 5 zeolite matrix with a main point of attention to potentially occurring CESR signals. Instead, only one paramagnetic signal is observed of symmetric shape at $g_c = 2.0025$ referring to C-based localized defects. The ESR signal features provide evidence for the occurrence of a low temperature ferromagnetic/spin-glass behaviour in the aligned arrays of sub-nanometer carbon nanotubes confined in zeolite nano-channels, owing to non-bonding localized states. Features related to the much anticipated conduction electron spin resonance could thus not be detected, which is tentatively ascribed either to excessive signal broadening or trapping of itinerant spins. Future work could attempt to observe ESR on a single crystalline particle of SAPO 5 with embedded SWCNTs rather than on powder, this in order to reveal any potential anisotropic effects, of much relevance for further insight.

<div align="center">***</div>

Several illuminating discussions with Dr. B. Do´ra are thankfully acknowledged.

**Figure Captions:**

**Fig. 1 (a):** First derivative K-band ESR spectra measured on SWCNT@SAPO 5 using $B_m$ = 0.3 G and $P_\mu$ = 1.3 nW. The signal at g ≈ 1.99876 stems from a co-mounted Si:P marker sample; **(b)** broad magnetic field range X-band ESR spectrum using 'enhanced' spectrometer settings ($B_m$ = 0.8 G, $P_\mu$ = 0.01 μW) to trace potential additionally present signals, without success however.

**Fig. 2:** The temperature dependence of the K-band ESR spectral parameters as obtained from optimized computer simulations indicating the signal to exhibit a highly Gaussian lineshape: (a) the peak-to-peak line width, (b) the g value, and (c) $\chi_{ESR}$ (∝ area under ESR absorption signal). The temperature dependence of $\Delta B_{PP}$ could be adequately described by $\Delta B_{pp}(G) = \Delta B_{pp}(\infty) + A\exp(-T/T_g)$ with $T_g$ = 7 ± 2 K, shown as a solid curve. In Fig. 2(b), the solid line marks the average g value. As evident from the inset in Fig. 2(c), the temperature dependence of the ESR susceptibility obeys a Curie-Weiss law over a broad T range (~ 15 – 90 K), from where a Curie-Weiss temperature $T_C$ = 4 ± 1 K is inferred through a least-square linear fitting (solid line).

**Fig. 3:** The microwave frequency f dependence of the observed peak-to-peak ESR signal width in confined SWCNT@SAPO 5 at 70 K. This reveals a weak dependence on f suggesting a small inhomogeneous (Gaussian) contribution to the line broadening. The solid line represents a least–square linear fit, described by $\Delta B_{PP}$ (G) = 7.7 + 0.018 f (GHz).

**Fig. 1:**

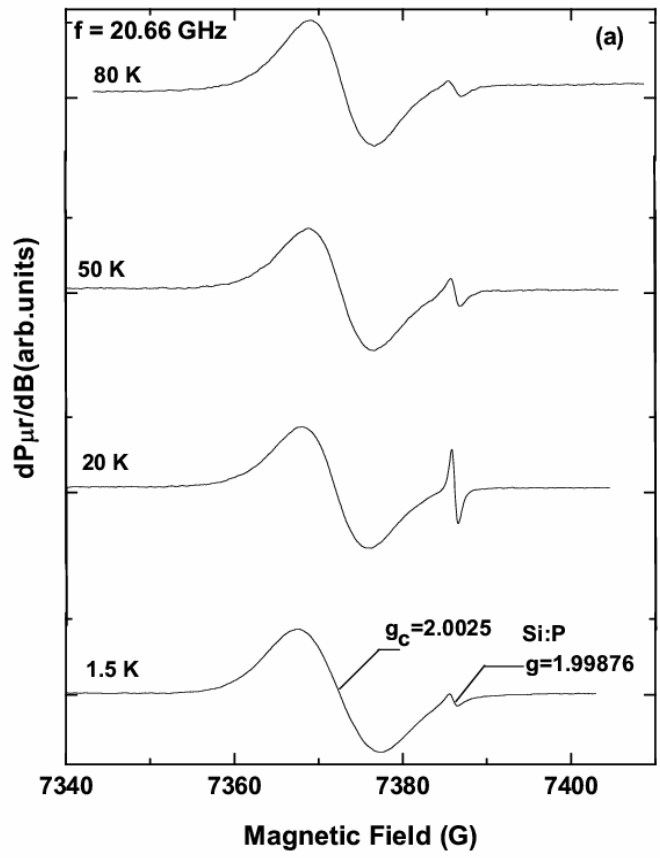

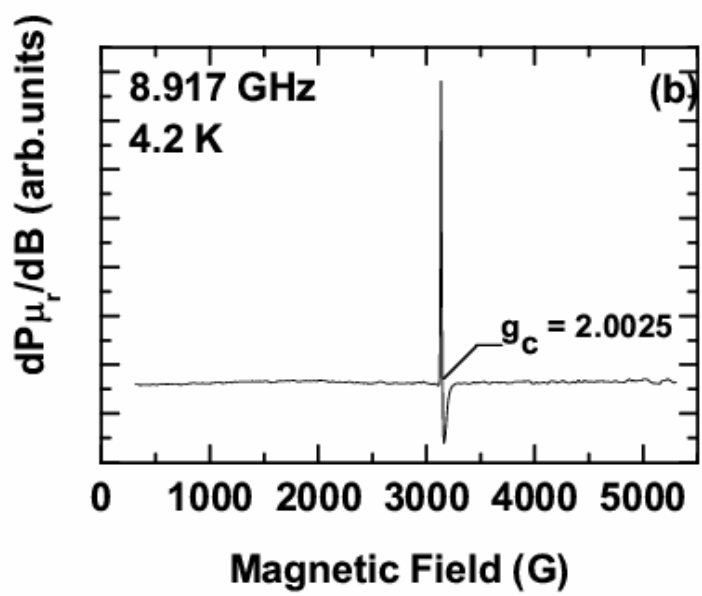

**Fig. 2:**

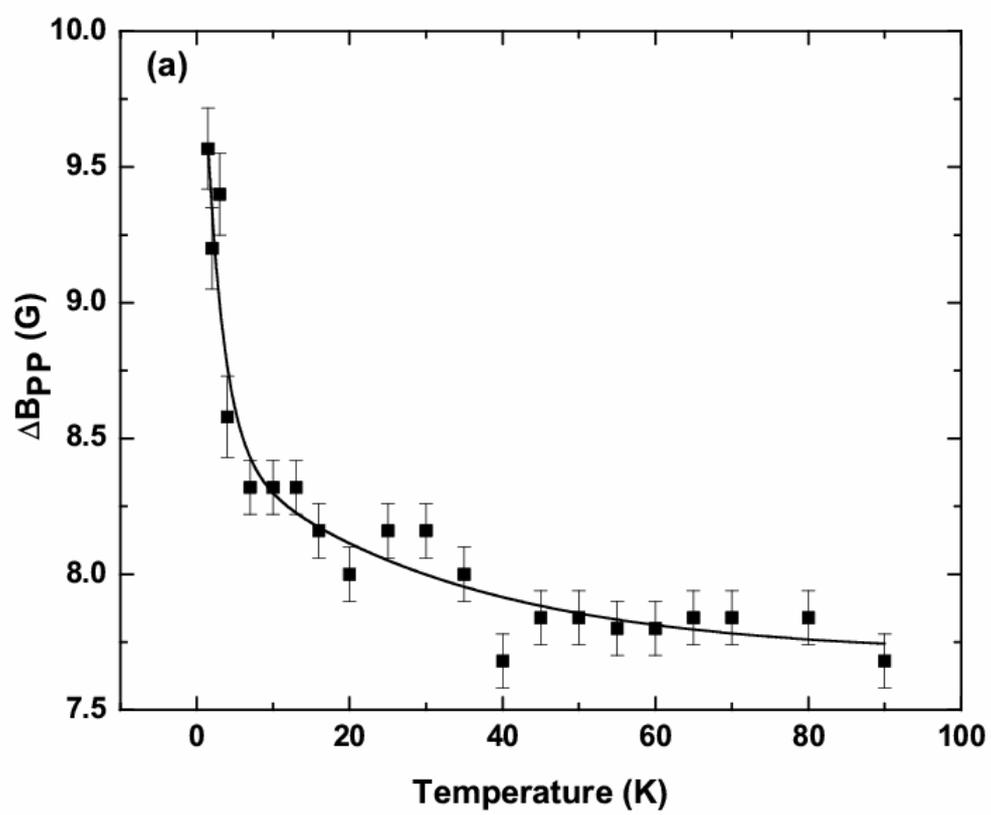

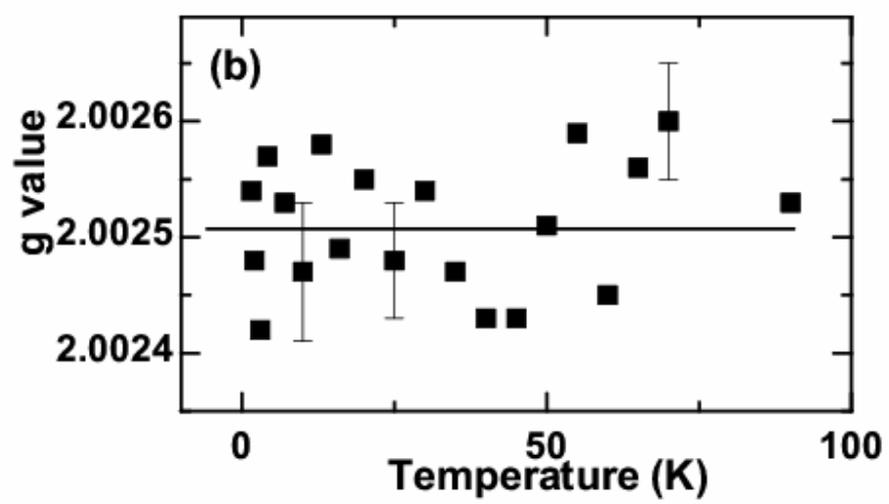

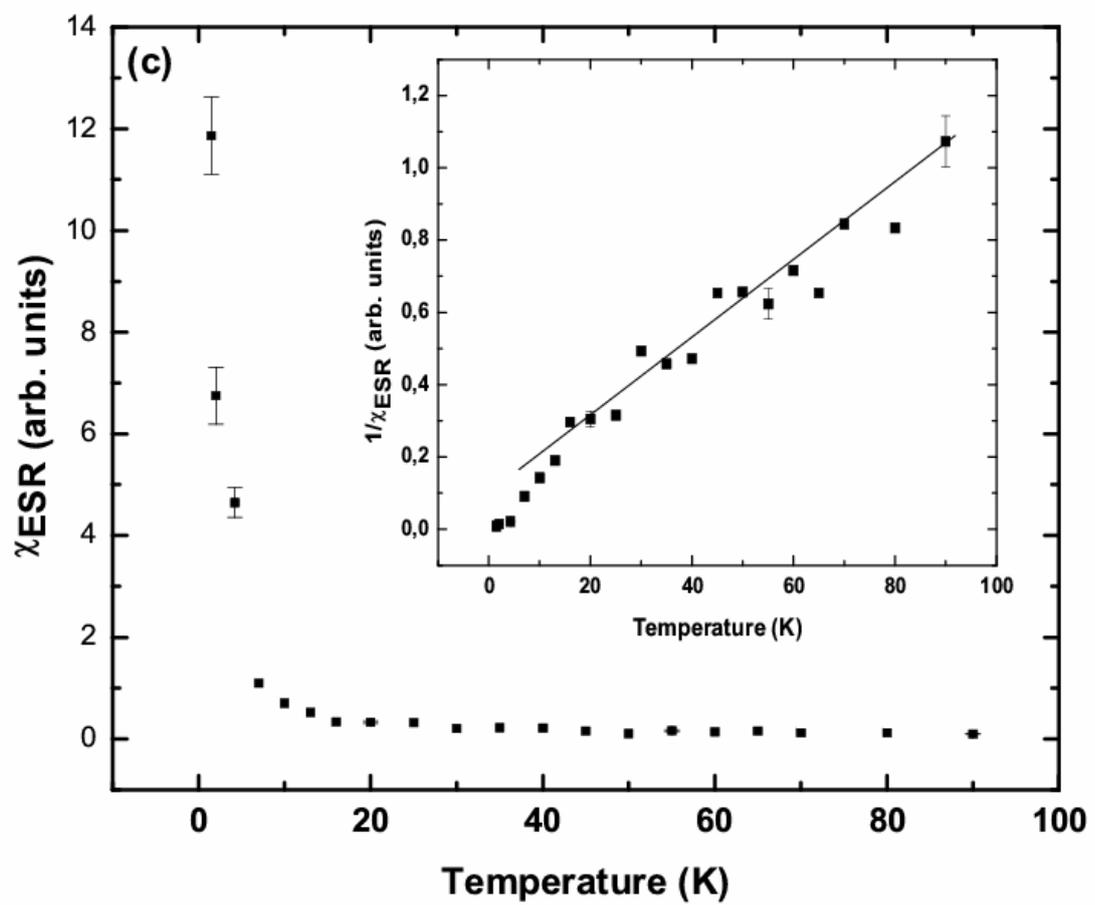

**Fig. 3:**

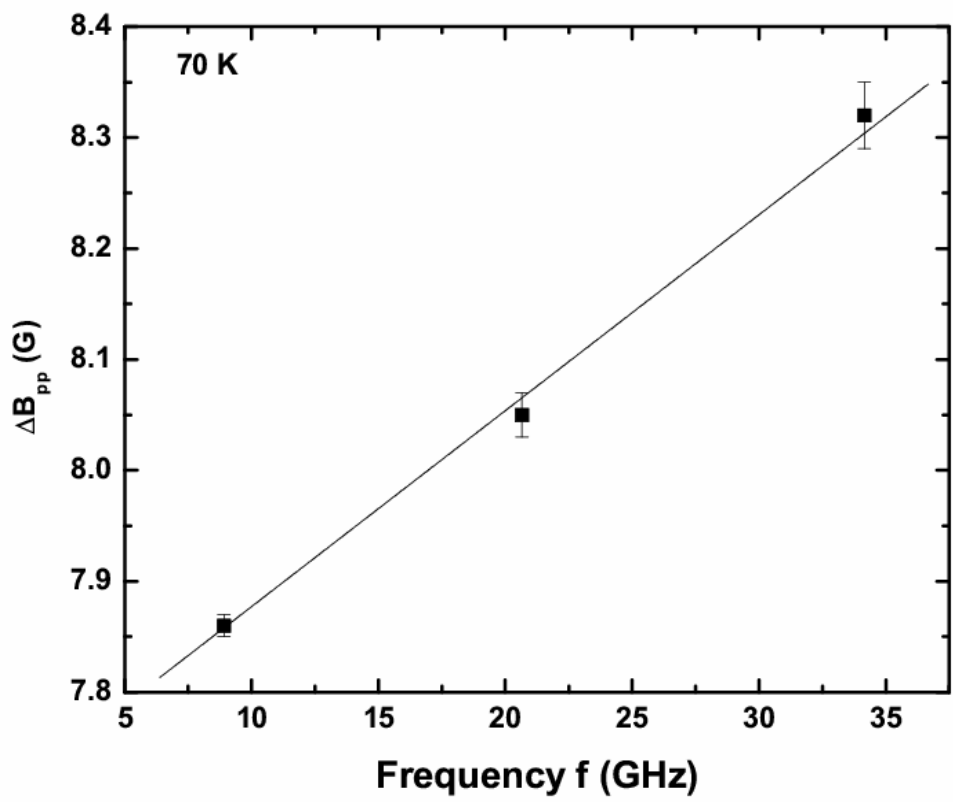